\documentclass[twocolumn,superscriptaddress,showpacs,pra]{revtex4}
\usepackage{epsfig}
\usepackage{epstopdf}
\usepackage{delarray}
\usepackage{amsmath, amssymb}
\usepackage{bm}
\usepackage{graphicx}
\usepackage{placeins}
\usepackage{color}

\begin{document}
\title{Self-Localized Solutions of the Nonlinear Quantum Harmonic Oscillator}

\author{Cihan Bay\i nd\i r}
\email{cihanbayindir@gmail.com}
\affiliation{Associate Professor, Engineering Faculty, \.{I}stanbul Technical University, 34467 Maslak, \.{I}stanbul, Turkey. \\
						 Adjunct Professor, Engineering Faculty, Bo\u{g}azi\c{c}i University, 34342 Bebek, \.{I}stanbul, Turkey. \\
						 International Collaboration Board Member, CERN, CH-1211 Geneva 23, Switzerland.}

\begin{abstract}
We analyze the existences, properties and stabilities of the self-localized solutions of the nonlinear quantum harmonic oscillator (NQHO) using spectral renormalization method (SRM). We show that self-localized single, dual and triple soliton solutions of the NQHO do exists, however only single and dual soliton solutions satisfy the necessary Vakhitov and Kolokolov slope condition, therefore triple soliton solution is found to be unstable, at least for the parameter ranges considered. Additionally, we investigate the stability characteristics of the single and dual soliton solutions using a split-step Fourier scheme. We show that single and dual soliton solutions are pulsating during time stepping. We discuss our findings and comment on our results.

\end{abstract}
\maketitle

\section{\label{sec:level1} Introduction}

Quantum harmonic oscillator (QHO) is one of the fundamental models in quantum mechanics \cite{Schrodinger, Griffiths, Liboff, Pauli, Messiah}. It can be viewed as the quantum mechanical analog of the simple harmonic oscillator of the classical vibration theory. The vibrations of atomic particles and molecules under the effect of restoring spring like potential due to molecular bonding are modeled within its frame \cite{Schrodinger, Griffiths, Liboff, Pauli, Messiah}. QHO admits exact solutions in terms of Hermite polynomials and can be extended to N-dimensions to model multidimensional atomic and molecular vibrations \cite{Schrodinger, Griffiths, Liboff, Pauli, Messiah}.

Nonlinear quantum mechanics studies, on the other hand, are becoming increasingly popular recently \cite{Nattermann, Reinisch, Wang_Nl_entang, Bay_Zeno, Bay_TWMS2017, Chia_Vedral}. Majority of the studies on nonlinear quantum phenomena are modeled in the frame of dynamic equations, i.e. the nonlinear Schr\"{o}dinger equation (NLSE). Compared to the linear Schr\"{o}dinger equation, NLSE can adequately model the cubic nonlinearity effects on the wavefunction. Such nonlinearities give rise to many interesting quantum mechanical phenomena including but are not limited to solitons, rogue waves, nonlinear quantum entanglement and quantum chaos. Analogs of these phenomena may also appear in the macroscopic physical environment.

Various studies which investigate the effects of nonlinearity on quantum oscillations do also exist in the literature \cite{Kivshar, Bay2019_qho_rogue, Carinena2007, Zheng, Ranada, SchulzeHalberg2012, SchulzeHalberg2013}. As discussed in the relevant literature, nonlinearity can arise in different ways. One possible way that gives rise to nonlinearity is the nonlinear behavior of bonding spring like stiffness and its corresponding potential. Another source of nonlinearity, which is investigated in this paper, arises due to strong electric and magnetic fields, which eventually leads to the cubic nonlinear term in the NQHO models.  

In this paper, we consider the NQHO model first proposed in \cite{Kivshar} and recently generalized by us \cite{Bay2019_qho_rogue}. This model equation can only be solved numerically, but for some limiting cases exact analytical solutions do exist \cite{Kivshar}. We study the self-localized solutions of this NQHO model using the spectral renormalization method (SRM). More specifically, we obtain the self-localized single, dual and triple soliton solutions of the NQHO using the SRM and discuss their properties. Additionally, we investigate the stability characteristics of those self-localized solutions using a split-step Fourier scheme which is used to perform the time stepping. We show that self-localized single and dual soliton solutions of the NQHO are stable and have pulsating behavior in time, at least for some of the parameter ranges considered in this paper. We also show that the self-localized triple soliton solution of the NQHO is not stable since it violates the necessary Vakhitov and Kolokolov slope condition for stability, at least for the parameter ranges considered.

\section{\label{sec:level2}A Nonlinear Quantum Harmonic Oscillator Model}

Linear quantum harmonic oscillator's (LQHO) Hamiltonian (energy) can be given by  \cite{Schrodinger, Griffiths, Liboff, Pauli, Messiah}
\begin{equation}
\widehat{H}=\frac{\widehat{p}^2 }{2 m}+\frac{1}{2}k x^2=\frac{\widehat{p}^2 }{2 m}+\frac{1}{2}m \omega^2 x^2
\label{eq01}
\end{equation}
In this formula $\widehat{H}$ denotes the Hamiltonian of the LQHO, $m$ denotes the particle mass and $k$ is the bonding stiffness of the atomic particle, which is analogous to spring constant in a classical mass-spring-dashpot system. In this formulation, the momentum operator can be given by $\widehat{p}=-i \hbar \partial / \partial x$  where $\hbar$ is the reduced Planck's constant. Thus, the unsteady Schr\"{o}dinger equation can be derived using the Hamiltonian formalism as
\begin{equation}
i \hbar \frac{\partial \psi}{\partial t}+\frac{\hbar ^2 }{2 m}\frac{\partial^2 \psi}{\partial x^2}-\frac{1}{2}m \omega^2 x^2 \psi=0
\label{eq02}
\end{equation}
where $i$ denotes the imaginary unity, $t$ is time variable, $x$ is the position and $\psi (x,t)$ is the wavefunction. This form of the LQHO is commonly studied in the literature \cite{Schrodinger, Griffiths, Liboff, Pauli, Messiah}. While majority of the studies on quantum harmonic oscillations utilizes linear models, few different forms of NQHO models are proposed in \cite{Kivshar, Bay2019_qho_rogue, Carinena2007, Zheng, Ranada, SchulzeHalberg2012, SchulzeHalberg2013}. Generally, two forms of nonlinearity are considered in these studies due to different nonlinear behaviors. One of them is the nonlinearly behaving molecular bond, which is commonly modeled using spring constant analogy, which is represented by different forms of the potential function than the commonly used trapping-well potential. However, in order to account for the effects of high-order electric and magnetic fields on wavefunction, a NQHO model is proposed in \cite{Kivshar}. The form of the NQHO proposed in \cite{Kivshar} can be given as
\begin{equation}
i\psi_t +  \frac{\partial^2 \psi}{\partial x^2}- x^2 \psi + \sigma \left|\psi \right|^2 \psi=0
\label{eq03}
\end{equation}
In here $\sigma$ is a constant which controls the strength of the nonlinearity. This equation can be derived by applying the non-dimensional parameters given in  \cite{Kivshar} to the LQHO and including the nonlinear term of the NLSE. Setting, $\sigma=0$, NQHO can be linearized and reduced to the LQHO, which admits solutions in the form of
\begin{equation}
\psi(x,t)=U(x) e^{-i \mu t}
\label{eq04}
\end{equation}
The analytical solution in this form can only be derived for the discrete spectrum of $\mu_n=1+2n$ where $n=0,1,2,...$ \cite{Kivshar}. For this discrete spectrum, the amplitude functions can be derived as $U_n=(2^n n! \sqrt{\pi} )^{-1/2} e^{-x^2/2}H_n(x)$ where $H_n(x)$ are the Hermite polynomials. These polynomials are described by
\begin{equation}
H_n(x)=(-1)^n e^{x^2/2} \frac{d^n (e^{-x^2/2})}{dx^n}
\label{eq05}
\end{equation}
giving $H_0=1$, $H_1=2x$, $H_2=4x^2-2$, $H_3=8x^3-12x$, ... etc. \cite{Kivshar, Abramowitz, Ryshik}. Eq.(\ref{eq03}) 
requires numerical solution except for some limiting cases \cite{Kivshar}, and a continuous frequency spectrum is considered in \cite{Kivshar} for its numerical solution.

In this paper, we study a slightly more general version of the NQHO given by Eq.(\ref{eq03}) which was first proposed by Kivshar \cite{Kivshar} and later extended by us recently \cite{Bay2019_qho_rogue}. To model the effects of variable potential due to varying bonding (spring) stiffness, we use a potential well constant, $\alpha$. Thus, the form of the non-dimensional NQHO equation studied in this paper can be written as
\begin{equation}
i\psi_t +  \frac{\partial^2 \psi}{\partial x^2}-\alpha x^2 \psi + \sigma \left|\psi \right|^2 \psi=0
\label{eq06}
\end{equation}
As before, the $t$ denotes the non-dimensional temporal parameter and $x$ denotes the non-dimensional spatial parameter. In the next sections of this paper, we study the existences and properties of the self-localized solutions of the NQHO given by Eq.(\ref{eq06}) using the spectral renormalization method (SRM). Additionally, using a split-step Fourier scheme, we study the stability characteristics of such solutions of the NQHO.

\section{Spectral Renormalization Method for Finding the Self-Localized Solutions of the Nonlinear Quantum Harmonic Oscillator}

There are few different techniques used to find the self-localized solutions of nonlinear systems. Some of these techniques are the shooting, self-consistency and relaxation techniques \cite{Petviashvili, Ablowitz, Yang, Fibich, Bay_CSRM}. One of the most commonly used methods for this purpose is the Petviashvili's method, in which the governing nonlinear equation is transformed into Fourier space similar to the other Fourier spectral methods using FFT routines \cite{Canuto, Karjadi2010,Karjadi2012, trefethen,  BayPRE1, BayPRE2, BayTWMS2016, Demiray2015, BayPLA, Baysci, BayTWMS2015, Bay_arxNoisyTun, Bay_arxNoisyTunKEE, Bay_arxEarlyDetectCS, BayMS}. Then, a convergence factor is used in accordance with the  degree of the nonlinear term \cite{Petviashvili, Ablowitz}. Petviashvili was the person who proposed this approach and he also applied this method to the Kadomtsev-Petviashvili equation \cite{Petviashvili}. In order to treat various forms of the homogeneities different than the fixed ones, Petviashvili's method is extended to the spectral renormalization method (SRM), which can be used to find the self-localized solutions of more general nonlinear equations \cite{Ablowitz,Yang, Fibich}. Later, another extension of the Petviashvili's method is proposed by us \cite{Bay_CSRM}, which is capable of finding the self-localized solutions in nonlinear waveguides under missing spectral information. This method is named as the compressive spectral renormalization method (CSRM) \cite{Bay_CSRM}. The SRM transforms the governing equation into wavenumber space using the Fourier transforms and then couples it to a nonlinear integral equation. The iterations are performed in the wavenumber space. Due to the coupling of these equations, the energy is conserved and the initial conditions converge to the self-localized solutions of the system studied \cite{Ablowitz}. Details and other possible uses of the SRM can be seen in \cite{Ablowitz}. 

In this section we apply the SRM to the NQHO model given by Eq.(\ref{eq06}) in order to study the self-localized solutions of the NQHO. We start with rewriting the NQHO model given by Eq.(\ref{eq06}) in the form of
 \begin{equation}
i\psi_t +  \psi_{xx}-V(x)\psi +\sigma \left| \psi \right|^2 \psi=0
\label{eq07}
\end{equation}
where $V(x)=\alpha x^2$ is the trapping-well potential function. Eq.~(\ref{eq07}) can be rewritten as
\begin{equation}
i\psi_t +  \psi_{xx} -V(x)\psi+ N(\left| \psi \right|^2) \psi =0
\label{eq08}
\end{equation}
where $N(\left| \psi \right|^2)=\sigma \left| \psi \right|^2$ is the nonlinear term of the NQHO. Using the ansatz, $\psi(x,t)=\eta(x,\mu) \textnormal{exp}(i\mu t)$, Eq.~(\ref{eq08}) becomes
\begin{equation}
-\mu \eta +  \eta_{xx} -V(x)\eta+ N(\left| \eta \right|^2) \eta =0
\label{eq09}
\end{equation}
where $\mu$ is the soliton eigenvalue. Iterations performed using the spectral representation of Eq.~(\ref{eq09}) may become singular \cite{Ablowitz}. In order to avoid the singularity of the scheme, a $p \eta$ term with $p>0$ can be added and substracted from Eq.~(\ref{eq09}) \cite{Ablowitz}. With this modification, the 1D Fourier transform of Eq.~(\ref{eq09}) can be found using the definition of the 1D Fourier transform as
\begin{equation}
\widehat{\eta} (k)=F[\eta(x)] = \int_{-\infty}^{+\infty} \eta(x) \exp[i(kx)]dx
\label{eq10}
\end{equation}
thus becomes
\begin{equation}
\widehat{\eta} (k)=\frac{(p+| \mu|)\widehat{\eta}}{p+\left| k \right|^2} -\frac{F[V \eta]-F \left[ N(\left| \eta \right|^2) \eta \right]}{p+\left| k \right|^2}
\label{eq11}
\end{equation}

The iteration formula given in Eq.~(\ref{eq11}) can be used to find the self-localized solutions of the NQHO model equation, however the iterations may diverge or it may tend to zero \cite{Ablowitz}. This problem can be solved by defining a new variable as $\eta(x)=\beta \xi(x)$, which has the Fourier transform given by $\widehat{\eta}(k)=\beta \widehat{\xi}(k)$. Using these substitutions, Eq.~(\ref{eq11}) can be rewritten as
\begin{equation}
\begin{split}
\widehat{\xi}_{j+1} (k) &=\frac{(p+| \mu|)}{p+\left| k \right|^2}\widehat{\xi_j}-\frac{F[V \xi_j]}{p+\left| k \right|^2} \\
& +\frac{1}{p+\left| k \right|^2}  F\left[\sigma { \left| \beta_j \right|^2 \left| \xi_j \right|^2} {\xi_j} \right] =R_{\beta_j}[\widehat{\xi}_j (k)]
\label{eq12}
\end{split}
\end{equation}
which is the iteration equation of the SRM in wavenumber domain. The algebraic condition on the parameter $\beta$, which prevents the scheme from diverging or tending to zero, can be derived using the energy conservation principle. Multiplying Eq.~(\ref{eq12}) with the $\widehat{\xi}^*(k)$ term, where $^*$ shows the complex conjugation, and integrating the resulting equation to evaluate the total energy of the system, it is possible to derive the algebraic condition as 
\begin{equation}
\int_{-\infty}^{+\infty} \left|\widehat{\xi} (k)\right|^2 dk= \int_{-\infty}^{+\infty} \widehat{\xi}^* (k) R_{\beta}[\widehat{\xi} (k)]dk  
\label{eq13}
\end{equation}
This becomes the normalization constraint, which prevents the scheme from diverging or tending to zero. The method summarized above is the SRM \cite{Ablowitz} and applied to NQHO in this paper. Starting the simulations using a single or multi-Gaussians as initial conditions, Eq.~(\ref{eq11}) and Eq.~(\ref{eq13}) are applied iteratively to find the self-localized solutions of the NQHO. The iterations are continued until the parameter $\beta$ convergences, for which the cut-off criteria is given for different simulations in the next part of this paper.  

\section{\label{sec:level3}Results and Discussion}
 
\begin{figure}[htb!]
\begin{center}
   \includegraphics[width=3.4in]{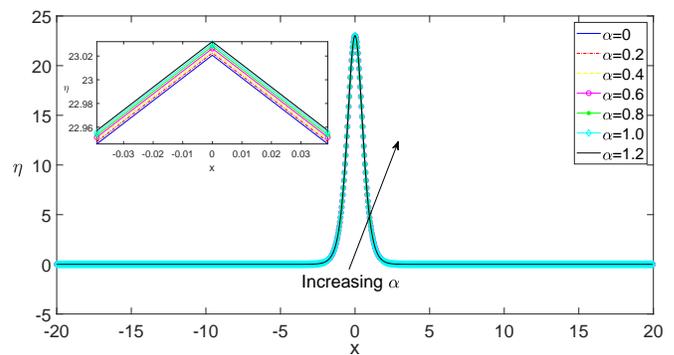}
  \end{center}
\caption{\small Self-localized single solitons for different trapping well potential strength, $\alpha$.}
  \label{fig1}
\end{figure}

In this section we apply the SRM summarized above to find the self-localized solutions of the NQHO model equation. In our simulations throughout this paper we use $N=1024$ spectral components. Starting with a single humped Gaussian in the form of $\exp{(-(x-x_0)^2)}$, where $x_0$ is taken as $0$, SRM converges to single humped self-localized soliton solutions of the NQHO. We plot these self-localized solutions of the NQHO model equation in Fig.~\ref{fig1} for various values of, $\alpha$, the trapping-well potential coefficient. Other parameters are selected as $p=30, \sigma=1, \mu=10.8$ for this simulation. The convergence criteria for this simulation is taken as the normalized change of $\beta$ to be less than $10^{-15}$. The single humped Gaussian converges to self-localized solution of the NQHO rapidly. The exact form of the analytical solution is unknown, however the profile resembles solitary waves in shape, however they have an asymmetric structure. This result can also be verified with the findings presented in \cite{Kivshar}. As one can realize from the figure, the increase in trapping-well potential strength results in bigger waves. One possible explanation for this result is that, bigger $\alpha$ values represents the trapping of the solitons in a more confined well, thus such bigger solitons are formed.

\begin{figure}[htb!]
\begin{center}
   \includegraphics[width=3.4in]{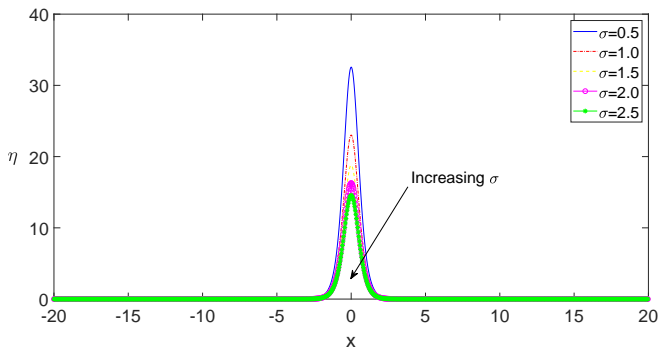}
  \end{center}
\caption{\small Self-localized single solitons for different nonlinearity strength, $\sigma$.}
  \label{fig2}
\end{figure}

Next we turn our attention to investigate the effects of nonlinearity coefficient, $\sigma$, on the existence and properties of the self-localized solitons of the NQHO. With this motivation, we depict Fig.~\ref{fig2}. As before, the simulation parameters are selected as $p=30, \mu=10.8$ and $\alpha=1$. Various values of $\sigma$ are used as indicated in figure and again, the cut-off criteria for the SRM iterations are selected as normalized change of $\beta$ being less than $10^{-15}$. As Fig.~\ref{fig2} confirms, self-localized solutions of the NQHO exists for various values of $\sigma$ as well and they tend to be smaller as $\sigma$ grows larger.

\begin{figure}[htb!]
\begin{center}
   \includegraphics[width=3.4in]{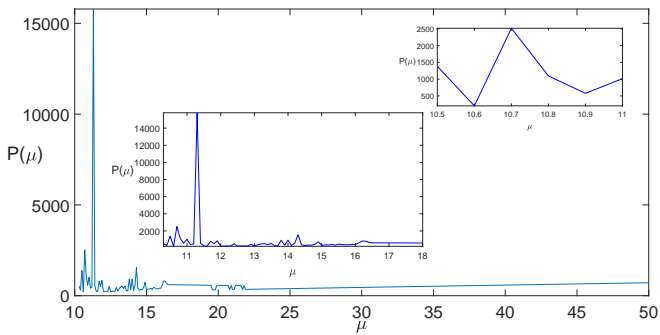}
  \end{center}
\caption{\small Self-localized single soliton power as a function of soliton eigenvalue, $\mu$.}
  \label{fig3}
\end{figure}

It is important to discuss the stability characteristics of the self-localized solutions of the NQHO model equation found using the SRM. For a soliton to be stable two conditions must be satisfied. The first conditions is known as the slope condition, $dP(\mu)/{d\mu}<0$, as first derived by Vakhitov and Kolokolov \cite{VakhitovStability, SivanStability}. In here $P=\int \left| \psi \right|^2 dx$ denotes the soliton power. Therefore the soliton power as a function of soliton eigenvalue must be examined to analyze the stability characteristics of the self-localized solutions of the NQHO.
With this aim, we plot the soliton power as a function of soliton eigenvalue in Fig.~\ref{fig3} for the single humped self-localized solution of the NQHO using parameters of $p=30, \alpha=1, \sigma=1$. We investigate the stability characteristics of such solitons within the soliton eigenvalue interval of $\mu=0-50$.

The figure clearly indicates that the  Vakhitov and Kolokolov slope condition, $dP(\mu)/{d\mu}<0$, is satisfied piecewise. This results suggests that self-localized solitons of the NQHO may be stable for some ranges of the soliton eigenvalue, $\mu$, such as $\mu \approx [10.7-10.8]$ for the parameters considered. Being a necessary condition for the soliton stability, the  Vakhitov and Kolokolov slope condition is not a sufficient one. The second condition for the soliton stability is the spectral condition. The spectral conditions states that the operator $L_+= -\Delta + V -1[N(\eta)-2\eta^2 N'(\eta)]-\mu$ for the NQHO problem that we analyze \cite{WeinsteinStability, SivanStability}, should have at most one eigenvalue which should be nonzero \cite{SivanStability}. In here, $\Delta$ shows the diffraction term, $V$ is the trapping-well potential and $N(\eta)$ is the nonlinear term given above. Spectral condition can be analyzed analytically or numerically. For various nonlinear models studied in the literature, a numerical approach is the more commonly used one.

\begin{figure}[htb!]
\begin{center}
   \includegraphics[width=3.4in]{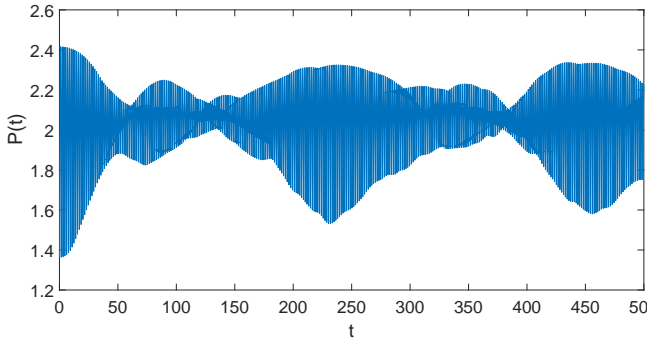}
  \end{center}
\caption{\small Self-localized single soliton power as a function of time, $t$.}
  \label{fig4}
\end{figure}

With this aim, we investigate the temporal stability of self-localized solutions of the NQHO model equation given by Eq.(\ref{eq06}) using a split-step Fourier method (SSFM). This SSFM is recently proposed by us \cite{Bay2019_qho_rogue} and it is validated using the analytical solutions of the some limiting cases of the NQHO model equation given by Eq.(\ref{eq06}). It is also tested against a $4^{th}$ order Runge-Kutta integrator. The SSFM, splits the governing NQHO model equation into linear and nonlinear parts. As a possible splitting, the nonlinear part can be written as
\begin{equation}
i\psi_t= -(-\alpha x^2 +  \sigma \left| \psi \right|^2)\psi
\label{eq15}
\end{equation}
which can be exactly solved. This integrations gives
\begin{equation}
\tilde{\psi}(x,t_0+\Delta t)=e^{i(-\alpha x^2 + \sigma \left| \psi_0 \right|^2)\Delta t}\ \psi_0   
\label{eq16}
\end{equation}
In here, $\psi_0=\psi(x,t_0)$ denotes the initial condition. $\Delta t$ indicates the time step, which is selected as $\Delta t=5\times 10^{-5}$ throughout this study, which does not cause stability problems. The remaining linear part of the NQHO model equation is
\begin{equation}
i\psi_t=-\psi_{xx}
\label{eq17}
\end{equation}
One can compute the linear part of the NQHO equation in periodic domain using spectral techniques. Using the most commonly utilized Fourier spectral technique, the linear part can be evaluated as
 \begin{equation}
\psi(x,t_0+\Delta t)=F^{-1} \left[e^{-i k^2\Delta t}F[\tilde{\psi}(x,t_0+\Delta t) ] \right]
\label{eq18}
\end{equation}
In here $k$ is the wavenumber parameter \cite{Bay_CSRM}. Inserting Eq.(\ref{eq16}) into Eq.(\ref{eq18}), the complete form of the SSFM iteration formula for the numerical solution of the NQHO model equation can be derived as
 \begin{equation}
\psi(x,t_0+\Delta t)= F^{-1} \left[e^{-ik^2\Delta t} F[ e^{i(-\alpha x^2+ \sigma \left| \psi_0 \right|^2 )\Delta t}\ \psi_0 ] \right]
\label{eq19}
\end{equation}
Starting from the initial conditions, we perform the integration of the NQHO model equation using this procedure. In order to study the temporal stabilities of the self-localized solutions of the NQHO model equations, the initial conditions are taken as the self-localized solutions found by the SRM, such as the ones depicted in Fig.~\ref{fig1}
and Fig.~\ref{fig2}. Using the normalized self-localized soliton obtained by SRM for $p=30, \alpha=1, \sigma=1, \mu=10.8$ as the initial condition, we observe a pulsating behavior during the time stepping. This can also be understood by checking Fig.~\ref{fig4}, where the soliton power as a function of time is depicted.

\begin{figure}[htb!]
\begin{center}
   \includegraphics[width=3.4in]{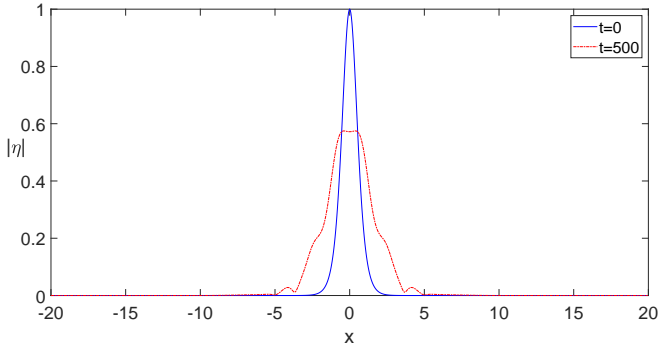}
  \end{center}
\caption{\small Self-localized single soliton at two different times, $t=0$ and $t=500$.}
  \label{fig5}
\end{figure}
In order to illustrate the pulsation properties of the self-localized solutions of the NQHO, we depict Fig.~\ref{fig5}. In this figure, the initial self-localized soliton profile and the same soliton after an integration time of $t=500$ is given. In our simulations we observe the pulsating recurrence type behavior between these two profiles, that is these forms are interchanging to each other gradually during time stepping in a recursive way.

\begin{figure}[htb!]
\begin{center}
   \includegraphics[width=3.4in]{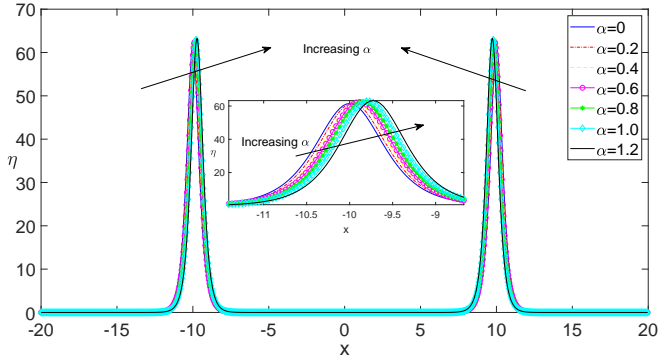}
  \end{center}
\caption{\small Self-localized dual solitons for different trapping well potential strength, $\alpha$.}
  \label{fig6}
\end{figure}
Next, we turn our attention to dual humped self-localized solitons of the NQHO model equation. Using $N=1024$ spectral components as before, selecting the computation parameters are $p=150, \sigma=1, \mu=10.8$, and specifying the convergence as the normalized change of $\beta$ to be less than $10^{-15}$ as before, we depict the dual humped self-localizes solutions of the NQHO model equation in Fig.~\ref{fig6} for various values of $\alpha$. The initial condition for the SRM is selected as $\exp{(-(x-x_0)^2)}+\exp{(-(x-x_1)^2)}$ for which the locations of the Gaussians are selected as $-x_0=x_1=10$.

\begin{figure}[htb!]
\begin{center}
   \includegraphics[width=3.4in]{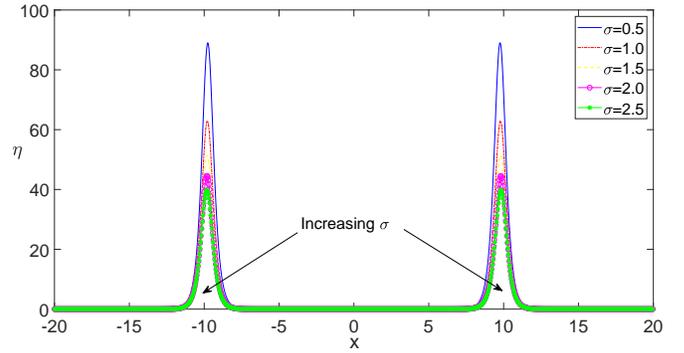}
  \end{center}
\caption{\small Self-localized dual solitons for different nonlinearity strength, $\sigma$.}
  \label{fig7}
\end{figure}
Setting $\alpha=1$ and keeping the other parameters as before, the dual humped self-localized solitons are obtained by the SRM for various values of the nonlinearity coefficient, $\sigma$, and they are depicted in Fig.~\ref{fig7}. As Fig.~\ref{fig6} and Fig.~\ref{fig7} confirms, the dual humped self-localized solutions of the NQHO model equation to also exist. As in the case of single humped solitons, the dual humped solitons are also asymmetric about the vertical axis, similar to the soliton profiles given in \cite{Kivshar}.

\begin{figure}[htb!]
\begin{center}
   \includegraphics[width=3.4in]{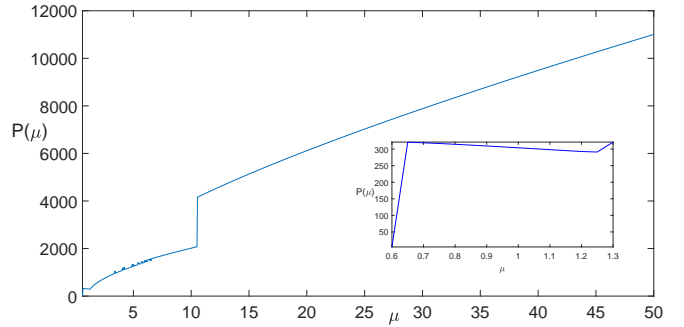}
  \end{center}
\caption{\small Self-localized dual soliton power as a function of soliton eigenvalue, $\mu$.}
  \label{fig8}
\end{figure}
In order to check the stability characteristics of dual humped self-localized solitons, we depict soliton power as a function of soliton eigenvalue, $\mu$, in Fig.~\ref{fig8}. The parameters of computation are selected as $p=150, \alpha=1, \sigma=1$ and the soliton eigenvalue interval of $\mu=0-50$ is scanned. As indicated in Fig.~\ref{fig8}, the Vakhitov and Kolokolov slope condition is only satisfied in a small interval of $\mu \approx 0.65-1.25$, thus stable solitons can be found in this range for the parameters considered.

\begin{figure}[htb!]
\begin{center}
   \includegraphics[width=3.4in]{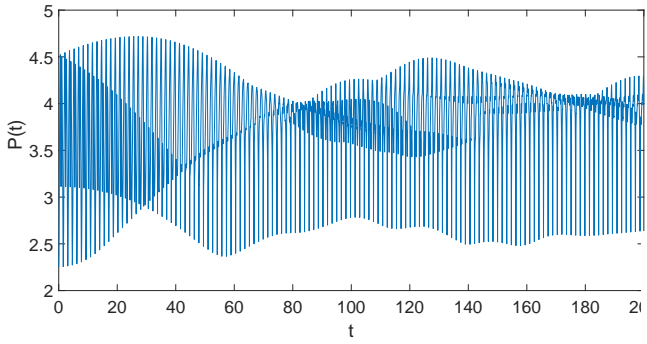}
  \end{center}
\caption{\small Self-localized dual soliton power as a function of time, $t$.}
  \label{fig9}
\end{figure}

\begin{figure}[htb!]
\begin{center}
   \includegraphics[width=3.4in]{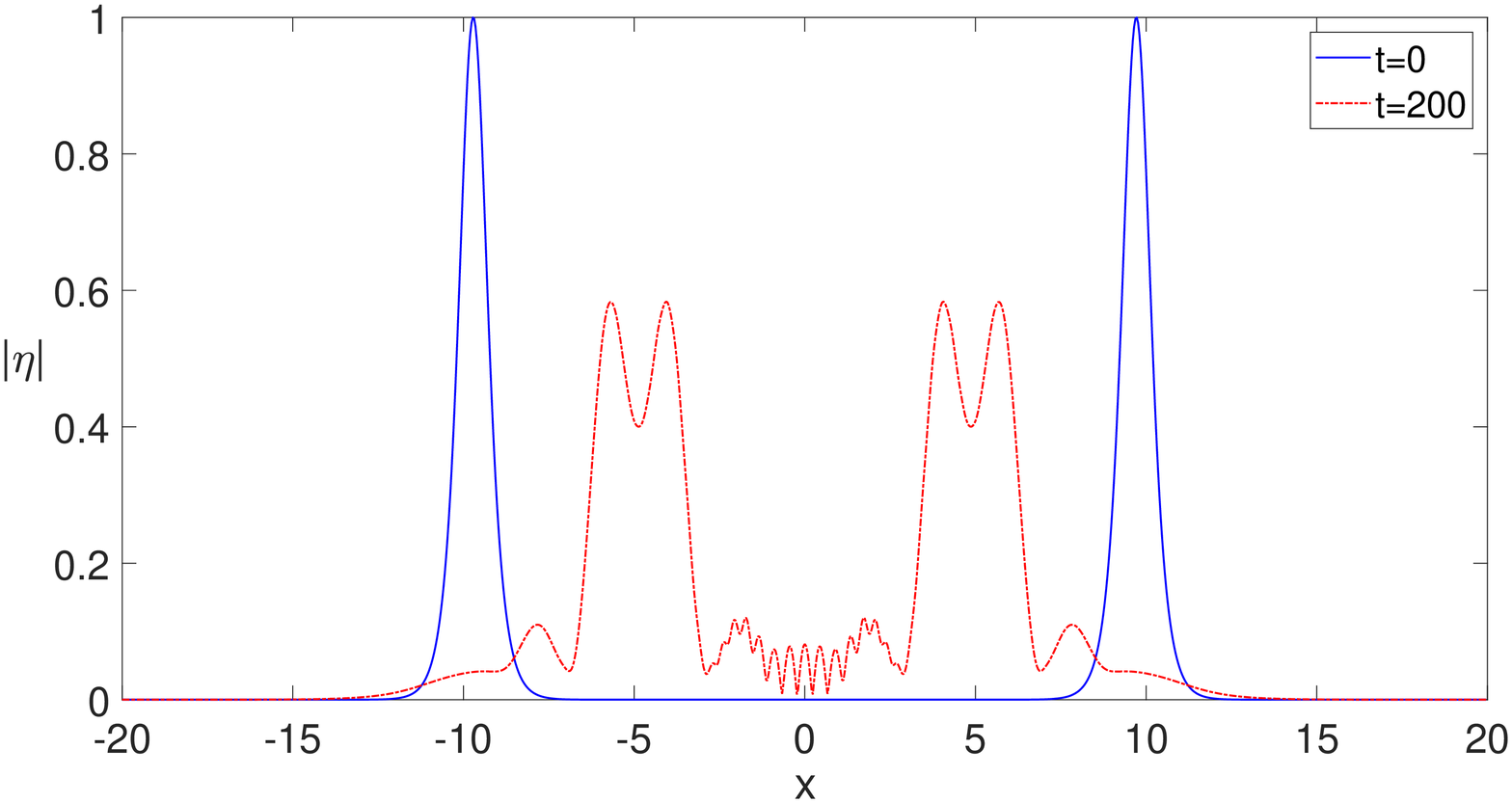}
  \end{center}
\caption{\small Self-localized dual soliton at two different times, $t=0$ and $t=200$.}
  \label{fig10}
\end{figure}
In order to check the temporal stability of the dual humped self-localized soliton we again use the SSFM summarized above. Starting the time stepping using the normalized dual humped self-localized soliton obtained for $p=150, \alpha=1,\sigma=1, \mu=1$, as the initial condition, the soliton power as a function of time is depicted in Fig.~\ref{fig9}, and the dual humped soliton at two different times of $t=0$ and $t=200$ is depicted in Fig.~\ref{fig10}. Similar to the single humped self-localized solitons, the dual humped soliton also exhibits a pulsating behavior, the form is gradually and recursively interchanging from the form given at $t=0$ to the one given at $t=200$.

\begin{figure}[htb!]
\begin{center}
   \includegraphics[width=3.4in]{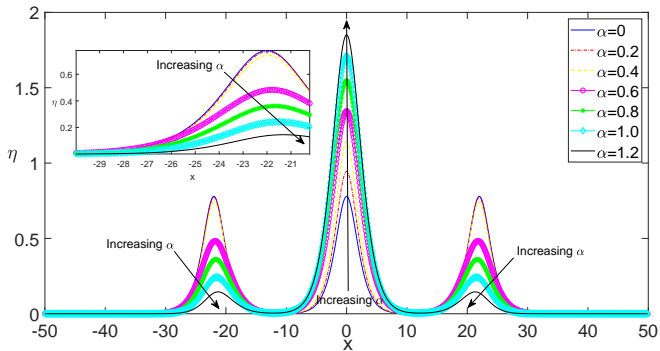}
  \end{center}
\caption{\small Self-localized triple solitons for different trapping well potential strength, $\alpha$.}
  \label{fig11}
\end{figure}
\begin{figure}[htb!]
\begin{center}
   \includegraphics[width=3.4in]{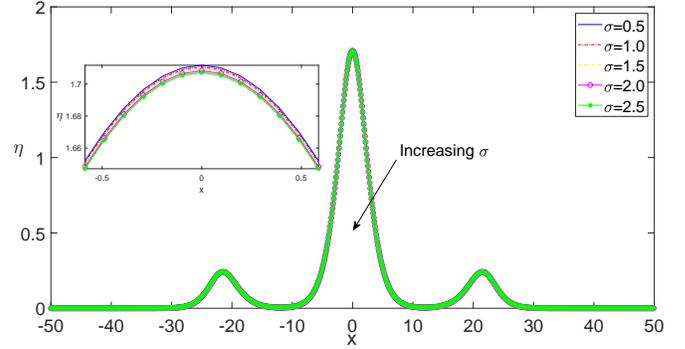}
  \end{center}
\caption{\small Self-localized triple solitons for different nonlinearity strength, $\sigma$.}
  \label{fig12}
\end{figure}

Lastly, we investigate the properties of triple humped self-localized solutions of the NQHO model equation obtained by the SRM. For the SRM to converge to such soliton solutions, parameters needed to be relaxed and are selected as $p=1000, \alpha=1, \sigma=1, \mu=1$. The convergence criteria for the SRM is also relaxed to be the normalized change of $\beta$ being less than $10^{-1}$. The initial condition used in SRM to obtain the triple humped solitons is selected as superimposed three Gaussians in the form of $\exp{(-(x-x_0)^2)}+\exp{(-(x-x_1)^2)}+\exp{(-(x-x_2)^2)}$ where $x_0=-10, x_1=0, x_2=10$. The triple humped self-localized solutions of the NQHO model equation are depicted in Fig.~\ref{fig11} and Fig.~\ref{fig12}, for various values of $\alpha$ and $\sigma$, respectively.

\begin{figure}[htb!]
\begin{center}
   \includegraphics[width=3.4in]{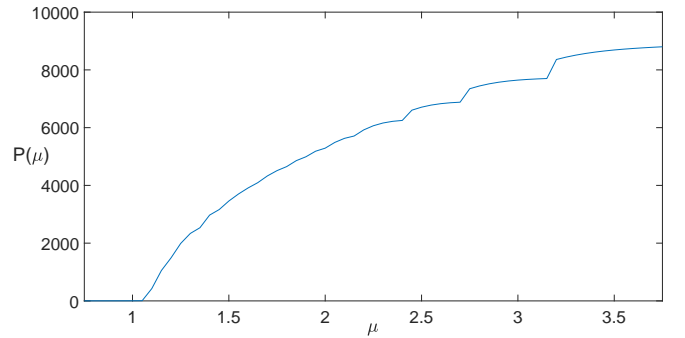}
  \end{center}
\caption{\small Self-localized dual soliton power as a function of soliton eigenvalue, $\mu$.}
  \label{fig13}
\end{figure}
Contrary to results presented for single and dual humped self-localized solitons of the NQHO model equation earlier in this paper, the soliton power vs soliton eigenvalue graph is depicted in Fig.~\ref{fig13} indicates that Vakhitov-Kolokolov slope condition is not satisfied for the triple humped self-localized solutions of the NQHO model equation. We should note that scanned range of the soliton eigenvalue is $\mu=0-50$, however only some part of it is depicted for illustrative purposes. Since the graph increases monotonically, it is possible to conclude that triple humped self-localized solitons of the NQHO model equation are unstable, at least for the parameter ranges considered in this study.

Findings and the computational approach based on the SRM and the SSFM we have proposed in this paper for the investigations of the self-localized solitons of the NQHO model equation can be used to analyze nonlinear quantum harmonic oscillations. Additionally atomic vibration resonances, bonding and bond breaking strength of molecules under the effect of nonlinear electric and magnetic fields and trapping well potentials having variable strengths can also be studied within this frame. The procedure proposed in this paper and our findings can also find many possible applications in the macroscopic level, such as Bose-Einstein condensation.

\section{\label{sec:level1}Conclusion and Future Work}
In this paper we analyzed the existences and properties of the self-localized solitons of a nonlinear quantum harmonic oscillator model. In order to do so, we implemented a numerical approach based on the spectral renormalization scheme and showed that single, dual and triple self-localized soliton solutions of NQHO do exists. We compared our findings with the results existing in the literature. We also studied stability characteristics of such soliton solutions of the nonlinear quantum harmonic oscillator model using a split-step Fourier scheme. We showed that single and dual humped self-localized solutions of nonlinear quantum harmonic oscillator model equation are stable within some subintervals of the soliton eigenvalue, which may be considered as a piecewise continuous spectrum. Time stepping analysis showed that single and dual humped self-localized solitons of the nonlinear quantum harmonic oscillator model equation are pulsating in time. However, the triple humped self-localized soliton solution of the nonlinear quantum harmonic oscillator turned out to be unstable since it violates the necessary Vakhitov-Kolokolov slope condition. Our findings can be used to analyze nonlinear quantum harmonic oscillations under varying molecular bond stiffness and/or nonlinear field effects. Some other similar phenomena observed at the macroscopic level, such as in the Bose-Einstein condensation, can also be investigated using our results and the framework proposed in this paper.

\end{document}